Citations versus expert opinions: Citation analysis of Featured Reviews of the American Mathematical Society

by Lawrence Smolinsky[1,4], Daniel S. Sage[1], Aaron J. Lercher[2], and Aaron Cao[3]

[1]Department of Mathematics, Louisiana State University, Baton Rouge, LA 70803 (USA)
[2]LSU Library, Louisiana State University, Baton Rouge, LA 70803 (USA)
[3]Carnegie Mellon University, Pittsburgh, PA 15213 (USA)
[4]Corresponding author: smolinsk@math.lsu.edu

## Abstract

Peer review and citation metrics are two means of gauging the value of scientific research, but the lack of publicly available peer review data makes the comparison of these methods difficult. Mathematics can serve as a useful laboratory for considering these questions because as an exact science, there is a narrow range of reasons for citations. In mathematics, virtually all published articles are post-publication reviewed by mathematicians in Mathematical Reviews (MathSciNet) and so the data set was essentially the Web of Science mathematics publications from 1993 to 2004. For a decade, especially important articles were singled out in Mathematical Reviews for featured reviews. In this study, we analyze the bibliometrics of elite articles selected by peer review and by citation count. We conclude that the two notions of significance described by being a featured review article and being highly cited are distinct. This indicates that peer review and citation counts give largely independent determinations of highly distinguished articles. We also consider whether hiring patterns of subfields and mathematicians' interest in subfields reflect subfields of featured review or highly cited articles. We re-examine data from two earlier studies in light of our methods for implications on the peer review/citation count relationship to a diversity of disciplines.

## Introduction

Two methods of evaluating the impact, quality, importance or other versions of value of a scientific work are peer assessment and informetric indicators. Peer assessment includes reviews of individual articles, reviewing for publication by referees and editors, reviewing for scholarly prizes and awards and honors, reviewing for grant support, and more (Lee, Sugimoto, Zhang, & Cronin, 2013). Peer reviewers ostensibly attempt to directly assess value, quality, and relevance. The meaning of citations is more ambiguous, but they have been used as indicators of value, impact, and even fame and pecuniary value (Cronin, 2005). Both citations and peer review are used as instruments of research evaluation. There is interest in comparing the two in terms of understanding both the significance of citations and the validity of citations in research evaluation.

Citation and publication networks cover nearly the entirety of academic literature. Counts of citations are available for articles indexed in the Web of Science (WOS), Scopus, Google Scholar, and, for mathematics, Mathematical Reviews (MR), available online as MathSciNet. The situation for peer reviewing is different. While the entirety of the literature indexed in Scopus and the WOS has undergone peer review from referees and editors, there is no systematic evaluation that allows comparisons of articles.

Although both peer review and citation analysis may reveal certain aspects of the value of scholarly work: importance, novelty, scientific usefulness, etc., it is not clear that they measure the same aspects of value. For example, Aksnes, Langfeldt, and Wouters (2019) conjecture that research quality has four independently varying qualitative dimensions, only one of which is significantly measured by citations. It is accordingly a question of central importance to understand the relationship between citation analysis and peer review, and indeed, there have been many studies on the subject. However, almost all such research has examined peer review of research groups, institutions, or individual scholars. Although most peer review takes place at the article level, Patterson and Harris (2009, p. 343) observe that there are "surprisingly few" studies at this level.

Mathematics deserves special attention in bibliometrics. We will discuss that mathematics—as an exact science—has narrower range of reasons for citing than in other fields. This makes citation analysis somewhat less complex in mathematics than in other disciplines. Accordingly, mathematics can serve as a useful laboratory for bibliometric investigations.

In mathematics, there is a collection of distinguished articles well-suited for exploring the relationship between peer review and citation analysis. Between 1993 and 2004, those articles and books deemed to be especially significant were selected to receive featured reviews in MR. Since the choices were made shortly after the articles appeared, they were made independently of citations. The main goal of the present study is to investigate consistency between these two measures of quality for mathematical research (citations and expert opinion) by concentrating on featured review articles in MR. If citation measures are becoming commonly used measures of quality, then is the meaning of quality changing?

Prestigious highly cited and featured review articles are not evenly distributed throughout all subfields of mathematics, and these distributions sheds light on the perceived importance of subfields. Two other phenomena related to the perceived importance of subfields are the hiring patterns in top mathematics departments and the interest of mathematicians. We explore the relationship between these various phenomena related to the perceived importance of subfields as measured by citations and experts.

## Peer Review

Peer review is used to assess various manifestations of scholarly work including reviewing submitted manuscripts and grant proposals, selecting prizes and awards, and evaluating research departments (Moed, 2005, p. 229-231). Peer review is paramount in scientific evaluation. Before an article can accumulate data on the WOS or SCOPUS, it must first pass peer review to be published. While non-peer reviewed information is widely available in the digital age and indexed on Google Scholar, a Sloan Foundation study surveyed 4,000 academic researchers and found that the influence of peer review is growing in the digital environment (Nicholas, Watkinson, Jamali, Herman, Tenopir, Volentine, Allard, & Levine 2015). However, one must also recognize that the methodology of peer review is not uniform.  For example, different journals can give very different instructions to their reviewers, and they can also make use of the gathered information in very different ways.

**Reliability**

In comparing measures of research quality, the reliability of the measures limits any potential correlations.  It is accordingly important to consider the reliability of peer review.  In particular, how strongly do the results of peer review depend on the choice of reviewers, the form of the review instructions, and the timing of the review?

Campanario's (1998) review of literature on peer review concluded that peer review is both high status and low reliability.  While reviewers are typically given instructions or guidance on evaluation criteria, Langfeldt (2001) in her study of grant peer review points out that reviewers interpret the criteria differently.  The situation is summed up by an oft-repeated pithy quote from a former co-ordinating editor of the Journal of the American Statistical Association: "All who routinely submit articles for publication realize the Monte Carlo nature of review" (Eysenck & Eysenck, 1992, p. 394).

Several studies on inter-rater reliability are discussed by Lee et al. (2013).  The studies by Bornmann and Daniel (2008b); Jackson, Srinivasan, Rea, Fletcher, and Kravitz (2011); Kravitz, Franks, Feldman, Gerrity, Byrne, and Tierney (2010); and Rothwell and Martyn (2000) primarily had kappa values below 0.15 with the largest outlier being 0.28.  These are all very low values (McHugh, 2012, Table 3), supporting the Monte Carlo nature of review.

In theory, a uniform method for peer review across an entire discipline might be used as a standard measure, but no such method exists in any field.  Perhaps the best approximation to a high peer review assessment is an article's acceptance—after review by referees and editors—in a well-respected subject-area journal.  In fields where there is a reasonable consensus on the hierarchy of journals, one can consider the prestige of the journal in which an article appears.  However, this is problematic, since journals are now commonly ranked using impact factors rankings (Wouters, 1999), not peer review.

Another source of unreliability for peer review comes from the potential for personal bias.  For example, some journals and grant organizations allow researchers to suggest or exclude potential reviewers. Coauthors are excluded in some fields but not others.  There may be elaborate restrictions on reviewers in a promotion case, including disallowing faculty members from any of the candidate's prior institutions. Most of these examples are to avoid positive bias, but positive bias for one individual may be negative for competitors.  See Lee et al. (2013) for a broad review of the literature on bias in peer review.

It may be reasonable to expect that peer review becomes more reliable when one focuses on the most distinguished articles. For example, whereas different evaluators might reach opposite conclusions about the publishability of a marginal manuscript, one might expect almost all referees to agree on outstanding work.  Since this study is restricted to Featured Review articles on MR, constituting less than 0.13% of all articles reviewed, peer review may be more reliable here than is typical from the discussed peer review study literature.  We could not find this issue investigated in the literature. We remark that such an investigation would need to avoid the use of citation metrics in ranking outstanding articles or journals.

## Citation Analysis

Citation counts of scholarly publications are widely used as a measure of research performance, and thereby as an instrument of research evaluation. In Moed's summary of important informetric indicators (2017, p. 51, Table 3.5), about half depend on the networks of citations and publications.[1] G. Nigel Gilbert began his influential article, "Some studies have used the number of citations received by a paper as an indication of its scientific quality, significance or 'worth'.[2] Likewise, the number of citations obtained by an author has been used to measure the impact of his or her work on the scientific community[3]" (1977, p. 114). More recently the National Research Council (NRC), which is the primary operating arm of the United States National Academies of science and engineering, reported that US faculty members were "generally in agreement that publications and citations were the most important factors in [graduate] program quality" (National Research Council, 2009, p. 12). Many bibliometrics researchers attempt to study citations and their meaning without believing they are necessarily a measure of value or impact. Others have endorsed it as a measure of value or impact.[4]

### Reliability and meaning

Whereas peer review is known to be unreliable, the notion of reliability does not even make sense for citation counts. Indeed, the citation count of an individual article is simply part of the historical record; it is open to analysis, but not to experimentation. A single article can be given to different scientists to be independently peer reviewed and compared. However, a single article does not admit independent citation counts. On the other hand, while the meaning of peer review is clear, this is not the case for citation counts. Individual referees can interpret review criteria differently, but at least specific review criteria exist. In contrast, the possible reasons for citing an article are much more amorphous. There are no set criteria required for making a citation, and an author's reason for including a particular citation may not be obvious.

The notion that citation counts reflect the impact or value of an article's contribution to science is attributed to Robert K. Merton's normative theory. Merton was a sociologist who has been recognized as the founder of the sociology of science. He also served on the advisory board on the Science Citation Index (Storer, 1973), which is now part of the WOS. In Merton's view, a citation "registers in the enduring archives the intellectual property of the acknowledged source by providing a pellet of peer recognition of the knowledge claim" (Merton, 1988, p. 622).

---

[1] Others are based on altmetric measures or peer review such as mentions on social media, patented based measures, grant funding, or prizes and awards.
[2] See Gilbert (1977) for references.
[3] See Gilbert (1977) for references.
[4] For example, Bornmann and Osório write, "we use citations as a measure of 'value', because citations are usually applied to assess the usefulness and the value of publications for other researchers (Bornmann, 2017)" (Bornmann and Osório 2019, p. 546).

Even if one accepts that citations are given for scientific utility or as recognition of scientific accomplishments, there are still complications and subtleties in understanding the meaning of citation counts. For example, Eugene Garfield considered the issues of negative citations, self-citations, methodological and review articles, journal prestige, and variation by discipline (Garfield, 1979). However, in his view, these issues did not justify the rejection of the normative theory as they could be overcome with appropriate methodological adjustments (Garfield, 1979, pp. 244-252). Garfield wrote, "…we know that citation rates say something about the contribution made by an individual's work, at least in terms of the utility and interest the rest of the scientific community finds in it" (p. 250). We remark that as the evaluation stakes heighten for researchers, new versions of these technical challenges arise, e.g., the formation of "citation circles" (Aksnes et al., 2019, p.7).

On the other hand, if one rejects the normative view of citations, then there is no simple way to summarize the meaning of citations, leaving their use in evaluation unclear. A citation may be a pellet of peer recognition, as Merton asserts, but the underlying reason for the peer recognition may have little to do with scientific utility. First, since the citer is not anonymous, the reference may be made out of self-interest. Second, there are no awarding standards for the citation other than perhaps being relevant in the eye of the author and/or editor. Perhaps it is naive to attribute an author's choice of references primarily to the Merton theory of recognizing scientific contributions and scientific utility rather than to a competing notion of economic utility, where authors choose their citations to achieve their goals of being read, respected, and recognized. This perspective is exemplified by G. Nigel Gilbert's (1977) article title, "Referencing as persuasion."

Blaise Cronin writes, "The Achilles' heel of citation is its residual subjectivity…" (2005, p. 169). If the failure to cite is probabilistic then the randomness may be studied and corrected or, perhaps naively, ignored as averaging out. MacRoberts and MacRoberts have long argued that the process is nonrandom and that scientists' citations are "highly biased": "The equation: cited=used, may be correct with many caveats, exceptions, corrections, and qualifications, but the equation: not cited=not used, is simply false" (2018, p. 476).

## Citation Analysis versus Peer Review

Do citations and peer review measure similar notions of impact or value? The question has been explored in studies comparing peer review assessments of academic programs, research groups, individual scholars, and articles. Surveys by Aksnes et al. (2019) and Bornmann and Daniel (2008b) and Blaise Cronin's book (2005) describe some of the studies. We will discuss some of the results most relevant to the present study.

In comparing two measures A and B—here peer review and citation counts—the reliability of A and B are relevant. With low or unknown reliability of A and B, more measurements of the correlation between A and B with non-overlapping data sets can help develop an understanding of the relationship between A and B. Various measurements are not replication studies since correlations between A and B will be a distribution rather than a number.

## Correlations

Before turning to particular articles that address the issue of comparing citations and peer review, we comment on correlations used to address the question. The interpretation of a correlation must be made in the context of the question posed. Suppose we have two instruments or indicators, A and B. If the qualities they measure have some common component, then one might expect a nonzero correlation, i.e., a statistically significant correlation. However, that does not mean the instruments substantively measure similar qualities.

### *Correlation as a measure*

In considering whether indicator A and indicator B are measuring the same quality or if A can replace B as a measure, then statistically significant correlations of 0.6 may be very weak. Consider an example from the first author's teaching. He tested math students with paper tests and computer-based tests in Calculus to see if the knowledge and skill measured were the same. Each student's test result was an ordered pair, (handwritten score, computer score). The correlation was r > 0.6 and was statistically significant. However, the scatterplot graph (Figure 1) makes it apparent that the notions of skill and knowledge measured by these computer and handwritten tests are different. There is a large variance in the abscissa and ordinate at each level. Both tests may measure some aspects of knowledge and skill, but the specific aspects seem different.

**Figure 1.**
*Scatterplot, bins, and quartiles for a data set*

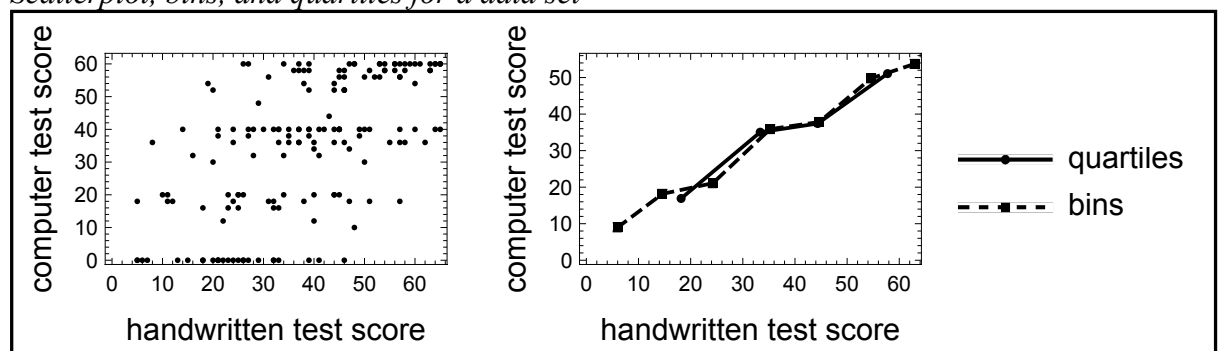

Data of individual students in scatterplot (left). The same data in bins of size ten and in quartiles.

### *Aggregating, averaging, and binning*

A second analytical tool that we feel requires caution is the use of averaging (or aggregating) data sets. In judging whether computer tests and handwritten tests measure the same aspects of knowledge, we would like to know if they are close on the level of individuals. Figure 1 may not give an appealing picture with a large variance at each level and suggest "no," but after averaging, the picture might suggest "yes". This issue also occurs if an indicator is too coarse with a small number of possible outcomes in one variable and averaging is done in the second variable. For instance, suppose we bin the abscissa on Figure 1 either by quartiles or in ten point groups and average on the ordinate for the binned groups. The Pearson correlations are then greater than 0.97, and the Spearman's rhos are both a perfect 1.

Binning, aggregating, and averaging may manifest in nonobvious ways. For example, peer reviewers might give a rating of 1-4 to approximate an unnamed underlying continuous rating. In citation analysis, one might make use of an impact factor that averages a large number of article citation results. We can now consider the main questions about the relation between peer review and citation counts. To what extent is the measure of value obtained using citations similar to the measure of value obtained using peer evaluations? More precisely:
1. Is there a statistically significant correlation between citations and peer review?
2. Do citations and peer review substantively measure a common notion?

As a caveat, we remark that a positive answer to the second question only makes sense if there is a high correlation between citations and peer review. However, the validity of even a high correlation between measures depends on the reliability (i.e., the self-correlation) of the measures, the second question with respect to the second question. As has already been discussed, reliability can be low for peer review and does not even make sense for citation counts. In light of this, we view a correlation of 0.6 as very weak for question 2.

## Studies

There are very few studies examining the correlation between citation counts and peer review at the article level. Patterson and Harris (2009) did one such study for articles in the journal *Physics in Medicine and Biology*. Patterson and Harris were an editorial board member and publisher, respectively, of this journal. They sought information on how to increase the impact factor of their journal and had access to internal peer review data. For the three years considered, they found statistically significant correlations between citation counts and peer review, all of which were weaker than 0.24. They used an averaging procedure where articles are aggregated into quintiles and then compared with the internal peer review. The authors thought it "reassuring to find that there is a significant correlation, albeit low, between citations and independent, expert, prospective review" (Patterson & Harris, 2009, p. 349). For editors interested in increasing an impact factor, this correlation may suffice to recommend action. However, this correlation, which is very low even after averaging, does not suggest that citation counts can serve as a reasonable replacement of the notion of value measured by peer review.

Other researchers have investigated the relationship between peer review and citations using data from F1000, a publisher of services for biological and medical scientists. F1000 does not provide systematic peer review, but rather is a form of social media for scientists allowing post-publication peer recommendation of articles. Recommendations are submitted by F1000 faculty members, who chose articles to read and recommend. Since only a small number of articles receive a recommendation, recommended articles can be usefully compared to highly cited articles. Two studies have included an examination of recommendations and WOS citations (Li & Thelwall 2012; Waltman & Costas, 2014). Both found weak but statistically significant correlations. Li and Thelwall used the ad hoc FFa numerical ratings provided by F1000 and Spearman's rho to find correlations of about 0.3. We discuss the larger study by Waltman and Costas in more detail. Of the 1,707,631 total publications in the total ("micro-subject" determined) population considered by Waltman and Costas, 38,327 had at least one recommendation and an assigned subject. They found that 73.7% of the highly cited (top 1%) articles have no recommendations. This information allows construction of the contingency table, Table 1. The correlation in Table 1 is $\phi = 0.163$ with a 95% confidence interval of [0.159, 0.168]. Given that there are less than half as many highly cited as recommended articles, the largest possible correlation was approximately 0.663, so that $\phi = 0.163$ is about 25% the maximum possible correlation.

**Table 1**
*Waltman and Costas's data*

| Highly Cited / Received a recommendation | Yes | No |
|---|---|---|
| Yes | 4,491 | 33,836 |
| No | 12,585 | 1,656,719 |

Author level studies are more common than the article level. If author level studies can be established as comparable to article level studies, then further progress on peer review/citation comparison is possible from author level investigations. Author level studies bin articles by authors. Binning independently selected articles would not change the correlations unlike the binning by ranks (e.g., quartiles) previously discussed. Articles binned by author are not independent but also not binned by ranks. A comparison with article level studies (our results and Waltman and Costas's results) are a preliminary measure of the validity of using author level studies.

Wainer and Vieira (2013) studied the relationship between bibliometric data and peer review coming from a Brazilian research funding agency. They looked at data for 2,663 individual scientists arranged in 96 groups by field and academic level. Their data can be plotted to give a distribution of correlations from their case study. They computed Spearman's rho correlations for each group and combined correlations from the same field using a weighted average method from biostatistics. Spearman's rho seems a minimal type of correlation to measure with citations, but Wainer and Vieira did not have direct peer review scores. They computed weighted Spearman's rho correlations for 55 fields (including humanities) between peer reviews and total citations for a researcher in each of WOS, Scopus, and Google Scholar resulting in 157 correlations (pp. 407-408, Table 3). We plot the distribution of correlations in Figure 2 and the mean of the correlations is 0.156, which is very small.

**Figure 2**
*Wainer and Vieira 157 correlation counts*

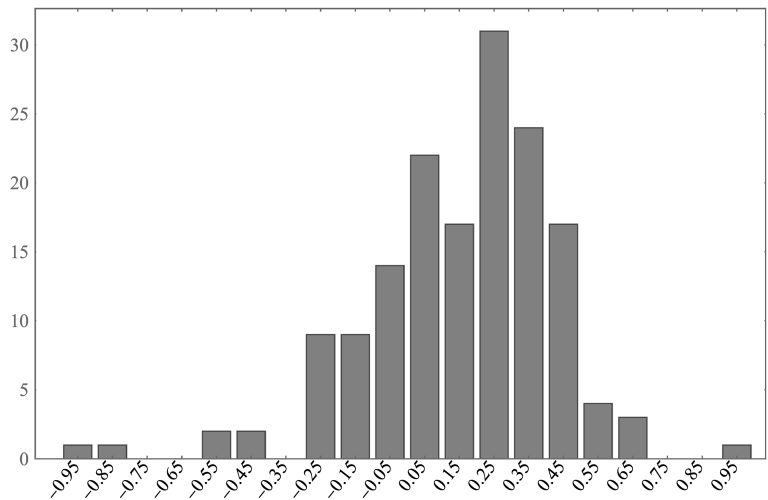

In summary, prior studies have found only low correlations between peer review and citation counts.  Moreover, data for precise, article-level comparisons is hard to come by.

## Mathematics and citation analysis

### Interest

One of the difficulties in citation analysis is the broad range of possible reasons for a given citation. The field of mathematics provides a useful test laboratory for understanding citations in general because in mathematics, this range is greatly restricted.  Mathematics has a standard of argument or proof that is not present in observational, experimental, or theoretical science.  Mathematical theorems are established by deductive reason from previously established results.  Accordingly, in the course of a proof, it is common for a mathematician to cite only articles containing lesser-known theorems used.  Refutation and debate of results become a small part of the literature.  As such, articles in mathematics tend to have fewer citations on average than is usual in science.

The point is illustrated by a conversation between the chemist, Darl McDaniel, and his mathematician son, Andrew McDaniel.  The mathematician described mathematical argumentation as a chain where each step is securely linked to the next in ironclad proof.  The chemist described argument in chemistry as a bundle of straw.  Here, each individual straw is a strand of evidence, with the strength of the argument determined by the number and thickness of the individual straws in the bundle.

Other than direct references to theorems used, the primary reason for citations in mathematics is to attempt to persuade readers of the interest, depth, and significance of the problems considered and results obtained. Table 2 is Bornmann and Daniel's (2008a) version of Eugene Garfield's (1962) list of possible motivations of citers.  We have added to it our view of its relevance to mathematics.

**Table 2**
*Motivations of citers*

| Reason for citation | Relevance in mathematics |
|---|---|
| 1.   Paying homage to pioneers. | Y |
| 2.   Giving credit for related work (homage to peers). | Y |
| 3.   Identifying methodology, equipment, etc. | Y |
| 4.   Providing background reading. | Y |
| 5.   Correcting one's own work. | N |
| 6.   Correcting the work of others. | N |
| 7.   Criticizing previous work. | N |
| 8.   Substantiating claims. | N |
| 9.   Alerting to forthcoming work. | Y |
| 10.  Providing leads to poorly disseminated, poorly indexed, or uncited work. | Y |
| 11.  Authenticating data and classes of fact (physical constants, etc.). | N |
| 12.  Identifying original publications in which an idea or concept was discussed. | Y |
| 13.  Identifying original publication or other work describing an eponymic concept or term | Y |
| 14.  Disclaiming work or ideas of others (negative claims). | N |
| 15.  Disputing priority claims of others (negative homage) (Garfield, 1962, p. 85). | Y |

In addition to its narrower uses of citations, mathematics has a lower average number of joint authors per article than other sciences (National Science Board, 2010, Table 5-16; Mallapaty 2018, with data provided by Larivière).  Moreover, since there is no laboratory work in mathematics, there are fewer collaborator or "team self-citations" (Garfield, 1979, p. 245). These facts simplify citation analysis for mathematics and, as was observed by Smolinsky, Lercher, and McDaniel (2015), may make it a closer fit to the preferential attachment model (Simon 1955;  Barabási & Albert 1999) or the cumulative advantage model (Price, 1976).

**Data**

MR (its online incarnation is MathSciNet) is a primary source for information on peer-reviewed articles and book series in the mathematical sciences.  MR is a searchable database of post-publication peer reviews, comprehensively covering the mathematical literature.  Published by the American Mathematical Society (AMS), over 125,000 new items are added each year (American Mathematical Society, 2019).  During the years 1995 to 2006, MR published 717,164 reviews of journal articles.[5]  For comparison, the WOS lists 163,648 articles that include mathematics as one of its categories for the publication years 1993-2004. Mathematics may be unique in having nearly its entire literature undergo post-publication review by scholars. Writing a review for MR is considered service to the profession similar to refereeing for a journal.  Since the reviewer is not anonymous, the reviewer has motivation to be diligent.  So MR is both comprehensive and authoritative for the mathematics profession.

---

[5] Article publication dates were 1993 to 2004.

MR does include a citation database, but the data is colored by peer review. Only certain MR indexed journals and book series are selected as "reference list journals" and only reference list journals may contribute references as citations. For example, the SCOPUS and MR indexed journal *Journal of Stochastic Analysis* is not a reference list journal. If an article in *Journal of Stochastic Analysis* is cited article by another article in *Journal of Stochastic Analysis*, then that citation is not credited in MR but is credited in SCOPUS. In addition, the coverage years for most journals in MR does not begin until 2000. For these reasons citation data from MR was not used.

From 1995 to 2006, MR recognized articles of particular note in *Featured Reviews*. Featured review articles were ". . . identified by the MR editors with the advice of distinguished outside mathematicians as being especially important…" (American Mathematical Society, 1995, p.1) and were highlighted on the title pages of MR and in MathSciNet. During the period 1995-2006, 927 articles were selected for featured review, constituting less than 0.13% of the MR literature[6] and less than 0.45% of WOS mathematics literature. The program was discontinued in 2006. The selection process was based on a posteriori peer review and was independent of citation counts, since the articles had already been accepted for publication or recently appeared.

In our determination that 927 articles received featured reviews, we made the following decisions. A few featured reviews include two articles that were published as complete articles (e.g., part 1 and part 2). Each of these articles is included in our count. Three other articles have corrections, entitled *Addendum ...*, *Correction ...*, or *Corrigendum ...,* that were separately published articles. These three are not included in our count. One article was published twice due to production errors in the original. We have counted the two versions as a single publication and added the three WOS citations to the original to the citation count for the corrected version. Among the 927 featured review articles, 79 are not indexed on the WOS and 734 include Mathematics as a WOS category. 80 featured review articles include a WOS classification of Applied Mathematics, 60 include one of the physics categories, and 30 include Mechanics.

We examined citation counts of featured review articles in bins of size 20 and size 5. The WOS lists 163,648 articles that include mathematics as one of its categories for the publication years 1993-2004. Usually, an article is termed highly cited if its citation count is in the top 1%. Here, this gives 1636 articles with 97 or more citations. However, in order to only consider full bins of 5, we restrict the definition to the 1559 articles with more than 100 citations. These are the top .952% most cited articles. All of the WOS highly cited articles are indexed in MR. The MR primary classification numbers were also recorded to examine the area distribution of the highly cited articles.

---

[6] The list of featured review articles is no longer available from the American Mathematical Society. We found featured review articles through the analysis of the review texts.

# Results

## Featured Review Articles Versus Highly Cited Articles

Of the 734 featured review articles that were indexed in the mathematics category on WOS, 122 were also highly cited. The correlation between the two dichotomous variables of being a featured review and being highly cited is the phi coefficient $\phi$, i.e., the mean square contingency coefficient. Three entries in the contingency table (Table 3) are available to compute $\phi$. The last necessary number in the contingency table is the number of articles $x$ that are neither a featured review nor a highly cited article. This last number would require knowing the number of articles in the intersection of the WOS mathematics category and the MR reviewed items, which was not computed. However, $0 \leq x \leq 163,648$,

$$\phi(x) = \frac{122x - 612 \cdot 1437}{\sqrt{(122 + 612)(122 + 1437)(x + 1437)(x + 612)}},$$

and $\phi(x)$ is an increasing function on $[0,\infty)$. For x > 9394, it is statistically significant at the 1% level using chi-squared (Chedzoy, 2006). The maximum possible value of the correlation is 0.11, but a correlation of $\phi = 0.11$ is weak. For x = 163,648, a 95% confidence interval is [0.091, 0.128]. We recognize that being highly cited is an artificial dichotomous variable, since it is determined by a cutoff value of the number of citations. We do not have enough information to conduct an exact point-biserial correlation calculation but estimate it to be less than 0.15.[7]

**Table 3**
*Contingency table for $\phi$*

| Highly cited / Featured review | Yes | No |
|---|---|---|
| Yes | 122 | 612 |
| No | 1437 | $x$ |

We note that since being a featured review is a rarer distinction (~0.45%) than being highly cited (~1%), there could not be a perfect correlation. Given the ex post facto rates of selection of featured review articles and highly cited articles, the largest possible $\phi$ would be 0.684. This is less than 16% of the possible maximum correlation. One can also consider Cohen's $\kappa$ statistic (Cohen, 1960), which has been previously used in the Italian study (Bertocchi, Gambardella, Jappellic, Nappi, & Peracchi, 2015) as well as for analyses of reliability. This statistic takes the observed categories' frequencies as an a priori given. For Table 3, $\kappa$ is

$$\frac{244 \, x - 1758888}{2293 \, x + 2689491}$$

---

[7] Using a sample of 6,000 and assuming that WOS mathematics category articles are included in MathSciNet.

and so $\kappa < 0.11$, which is small.

Only 7.83% of the 1559 highly cited WOS mathematics articles were featured review articles and only 16.62% of the 734 featured review articles classified in the WOS mathematics category were highly cited.  In Figure 3, the highly cited featured review articles represent only the tail of the distribution while the first 5 bars represent the 83.38% of featured review articles that are not highly cited.

**Figure 3**
*WOS citations versus number of featured reviews*

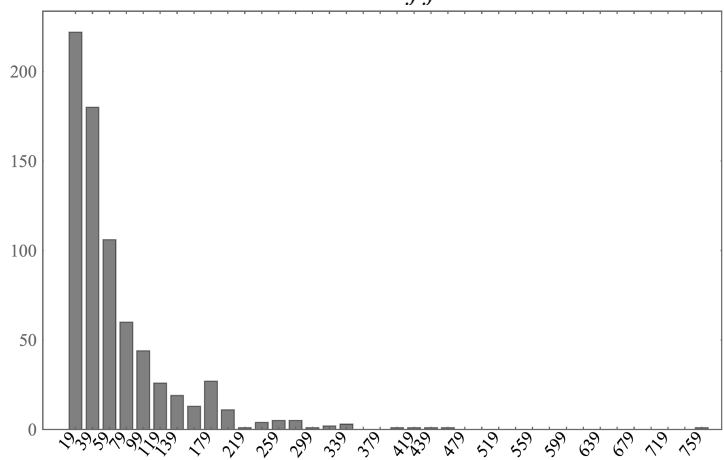

Frequency of WOS core citations for the 734 featured reviews indexed in the mathematics category on the WOS.

To summarize, the two notions of significance described by being a featured review article and being highly cited are substantially distinct.  This indicates that peer review and citation counts give largely independent determinations of highly distinguished articles—at least when peer judgment is uninfluenced by knowledge of citation counts.

**Subfield Analysis**

Data on featured review articles can also be used to investigate how subfields of mathematics are evaluated for their importance to mathematics as a whole.  How do the subfields of a discipline relate to hiring patterns and faculty interest?  Do the subfields chosen for hiring by distinguished departments correlate more strongly with the subfields with a larger number of highly cited articles or with those with more featured review articles?

The Mathematics Subject Classification (MSC) used by MR divides mathematics into 63 major topics. The Joint Data Committee of the American Mathematical Society, American Statistical Association, the Mathematical Association of America, and the Society for Industrial and Applied Mathematics[8] has aggregated the 63 topics into twelve "field of thesis" categories.

---

[8] Formerly, the committee also included a representative of the Institute for Mathematical Statistics.

Following the approach of Smolinsky and Lercher in their study of the effect of subdiscipline on citation rates (Smolinsky & Lercher, 2012), we will view these categories as the subfields of mathematics. Here, we consider two measures of the prominence of a subfield within mathematics. First, we will look at the subfield of interest of mathematicians. The professional mathematical societies request that members select two-digit MSC numbers as their fields of interest. The AMS generously supplied the 2009 data for the research of Smolinsky and Lercher (2012). Second, we examine the subfields of new PhDs hired from 2000–2010 by the top 48 mathematics departments (American Mathematical Society Group 1).

Let *FR*, *HC*, *H*, and *AMS* be real-valued random variables with domain the set of twelve fields {Algebra, Analysis, Geometry, Discrete, Probability, Statistics, Applied, Computation, Control, Differential Equations, Math Education, Other}. The random variables are defined by *FR*(field) = the number of featured review articles in the field, *HC*(field) = the number of highly cited articles in the field, *H*(field) = the number of Group 1 hires in the field as detailed in Smolinsky and Lercher (2012), and *AMS*(field)= the number of AMS members with responses indicating primary interest in the subfield. The correlation matrix for the random variables is given in Table 4. The correlation between subfield of hiring in the top departments and the featured review article subfields was very strong. It was still strong, but less so, between subfield of hiring and the subfield of highly cited articles. It is also noticeable that the subfields of faculty interest correlate more strongly with featured review article subfields than with the subfields of highly cited articles or hiring. All of the correlations in Table 4 between the random variables are statistically significant.

**Table 4**
*Correlation matrix*

| r.v. \ r.v. | *FR* | *HC* | *H* | *AMS* |
|---|---|---|---|---|
| *FR* | 1 | 0.71 | 0.91 | 0.89 |
| *HC* | | 1 | 0.80 | 0.67 |
| *H* | | | 1 | 0.77 |
| *AMS* | | | | 1 |

r.v. = random variable

The usage of either peer review or citation counts for recognizing those articles of particular distinction is subject to subfield biases. We observe that the subfields of featured review articles (*FR*) reflect the peer preference for the subfields as measured both by faculty interest (*AMS*) and hiring (*H*).

## Discussion

In this study, we examined the relationship between peer review and citation counts in mathematics by focusing on a body of highly distinguished mathematical articles, those selected for featured reviews and those highly cited. We find the relationship between peer selection and being highly cited is negligible. In fact, the phi correlation of 0.11 has a specific interpretation just as the Pearson r correlation does (McHugh, 2018). It is likely there is a relationship between peer selection and being highly cited because the phi value is statistically significant, however that relationship is negligible with less than 2% shared variance. The 2% overlap between being selected as a featured review and as highly cited is negligible.

A disconnect between assessment by peer review and citation counts in mathematics is visible from within the journal hierarchy, which reflects the journal acceptance and reviewing requirements. There is a higher level of consensus among mathematicians on which are the important mathematics journals than bibliometric measures indicate (Bensman, Smolinsky, & Pudovkin, 2010). Within the mathematical community, the standards of peer review for *Annals of Mathematics* are recognized as the most demanding. But these high standards are not reflected in correlation with high citation counts. In terms of being highly cited, *Annals of Mathematics* did more poorly than featured review articles.

Compare our result with the other article-level large study discussed—the F1000 biology study by Waltman and Costa (2014). Waltman and Costa's F1000 recommendations are social media selection and are less systematic than featured reviews, which covers all mathematical literature using assigned reviewers. We expected that a featured review selection would be a more reliable method of detecting the relationship between elite peer review and high citation counts. However, that was not the case and Waltman and Costa's result is consistent with ours. We computed phi for Waltman and Costas's data (2014) in Table 2 and it results in a shared variance between being highly cited and receiving a recommendation of less than 3%. The teaching illustration we gave in the left-hand side of Figure 1 was a shared variance of over 36%. The shared variance of less than 3% for peer selection and citation selection in mathematics and biology indicate there is little in common between peer and citation selection. Waltman and Costas also examined if the relationship was stronger by weighing the recommendation by the number of recommendations in their Table 2 (2014, p. 441). The correlation is a Pearson correlation of 0.27, which is somewhat better, but still only a shared variance of 7.3%.

Wainer and Vieira (2013) studied a diverse set of disciplines resulting in a descriptive analysis suggestive of the peer review/citation count relationship for articles. Wainer and Vieira's population sample is a convenience sample rather than random and so the study gives a descriptive statistical analysis rather than inferential—a common situation in bibliometric studies. Nevertheless, the Wainer and Vieira study gives a bell-like histogram for correlations with a mean 0.156. One expects the correlations to reflect the peer review/citation count relationship for articles if the binning articles by author were independent and to be somewhat high if not. Author-based binning may give upper bound estimates to the peer review/citation count relationship for articles. The low mean suggests points to conjecturing weak upper bound to this relationship across disciplines. Note that we only say weak because Spearman's rho correlations do not lend themselves to the same interpretation as phi or Pearson's r.

The present study points to a negligible peer review/citation count relationship and examining Waltman and Costas's altmetic study suggests a negligible relationship in biological sciences. Wainer and Vieira's study points to conjecturing a weak upper bound across disciplines.

Value and impact are judgments made by people. It may be incorrect to anthropomorphize citation counts as rendering such judgment—particularly when their meaning is unclear. We find sociological and philosophical contemplation (such as Table 2) to be thought-provoking in possibly forming hypotheses, but insufficient to render judgment on research. The Scientific Citation index was conceived as an aid to researchers to find connected research. It was inspired by a legal index to court cases. (Garfield, 1979). It was not designed to measure value or impact. Perhaps it is reasonable that imposing its use as a tool to measure value and impact gives results at odds with experts' opinions. The disconnect between high peer measures and high citation counts points away from value and impact as a description for citation measures. In the Merton view, impact may be a reasonable description of measurement by citations, but utility (to authors) may be more apt and conforming to Gilbert.

It appears that peer review and citation metrics are related to different notions of value in an article. Li and Thelwall suggest, F1000 evaluators measure "the quality of articles from an expert point of view, citations measure research impact from an author point of view…" (2012, p. 549). We believe peer review in general can be characterized as measuring quality from the expert point of view. But what does the expert or author point of view mean?

Peer review is a serious professional responsibility. It is a matter of basic professional ethics to be impartial and to review an article, researcher, program or institution according to the specified parameters without personal bias. The underlying assumption is that reviewers will embrace this responsibility and will not violate the trust of the profession to chase a (typically small) measure of personal career gain. In those cases where there is a significant conflict of interest, scholars are expected to recuse themselves. In peer review, the reviewer is functioning as an independent expert.

Since scholarly output is the basis of an academic's career, an author necessarily has a different viewpoint from that of an independent expert. An author is a consumer of references and a producer of articles. As producers, authors want their articles to be read, cited, and recognized as significant. As consumers of references, they will be guided by the economic utility of achieving their career goals. Consider the eight positive "relevant to mathematics" reasons for citation in Table 2. Other than results in the immediate chain of logical argument (item 3), there is great flexibility for an author to choose references for their economic utility. Which articles should an author include and exclude as "relevant?" Will the citation affect the likely peer reviewers? Will the citation increase the credibility of the article or attract readers and citers? Will citing fashionable or important articles improve the perception of importance? Suitable results may occur in multiple articles.

In mathematics, it is easy to see how an article could be flagged as significant by reviewers even though it is predictable that it will not be highly cited. One example is an article that solves a long-explored problem and completes a line of investigation. The solution may not open new directions of research, and even if it does, those new directions may not be of particular interest to present researchers. The article may not garner many citations since relatively few articles build on it.

Hiring in top departments as well as the list of fields of interest to mathematicians are more closely correlated with featured review subfields then with highly cited article subfields. It may be that the faculty, hiring committees, and chairs are acting as experts (reflecting peer review) when making hiring decisions. On the other hand, it is reasonable that selections of featured review articles would follow the subject pattern of the discipline members' interests.

There is a trend of viewing citation counts as the primary measure of the value or impact of an article. This disconnect between high peer measures and high citation counts may amount to a shift in the very meaning of value and impact used in describing academic articles. An illustration of is the 2010 NRC's assessment of USA research doctorate programs (National Research Council, 2011). While the quality of research faculty is important in the education of PhD scientists, mathematicians, engineers, and scholars, the NRC did not use direct expert opinion this assessment. The NRC only used expert opinion to weigh the importance of various data sets (National Research Council, 2009). Experts could express the relative weight of cites per publication compared to publications per faculty member or percent of female faculty, but not whether cites per publication was an adequate measurement of quality. It was further confounded by the use of co-author weighing (Smolinsky & Lercher, 2020). It is an influential study that likely has consequences for institutions and whole disciplines (Smolinsky & Lercher, 2020).

There are methodological limitations on studies comparing peer review and bibliometrics. Such studies usually involve data gathered for other purposes and so do not follow experimental protocols or journal peer-review protocols. Common issues are: a) reviewers are not assigned but self-selected, b) articles reviewed are not assigned but reviewer-selected, c) reviewers are not anonymous, d) reviewers have access to citation information, and e) reviewers know the journal where the article was accepted. Since single-blind review is the most common protocol in the sciences, we have omitted the lack of anonymity of authors from this list. However, Tomkins, Zhang, and Heavlin (2017) found articles with famous authors or from high-prestige institutions are at an advantage in single-blind review compared to double-blind review. Three studies at the article level are considered in this article: Patterson and Harris (2009), Waltman and Costas (2014), and the present study. Patterson and Harris does not suffer from any of these issues, all but d are relevant for Waltman and Costas, and c and e are present in the current study.

The American Mathematical Society also provides a demonstration of the acceptance of citations as the measure of value. Not only did the American Mathematical Society terminate the featured review program, but when the first author requested the list of featured review articles from the AMS, he was told that it was no longer available. Instead, he was offered the list of highly cited articles. We feel that this trend is unfortunate and identifying important articles from the viewpoint of independent experts is valuable to the community of scholars.